\documentclass[10pt]{article}

\usepackage[T1]{fontenc}
\usepackage{lmodern}
\usepackage[a4paper,top=20mm,bottom=25mm,left=17mm,right=17mm]{geometry}
\usepackage{graphicx}
\usepackage{amsmath}
\usepackage{amssymb}
\usepackage{booktabs}
\usepackage{array}
\usepackage{tabularx}
\usepackage{multirow}
\usepackage{etoolbox}
\usepackage{placeins}
\usepackage[authoryear,round]{natbib}
\usepackage[font=small,labelfont=bf]{caption}
\usepackage[hidelinks]{hyperref}
\usepackage{xurl}

\newcolumntype{Y}{>{\centering\arraybackslash}X}
\AtBeginEnvironment{thebibliography}{\small\raggedright\setlength{\itemsep}{1pt}\setlength{\parskip}{0pt}\Urlmuskip=0mu plus 1mu\relax}
\title{PriSAR: 3D Geometric-Prior-Guided Diffusion for Parameter-Controlled SAR Image Generation}
\author{
Fan Zhang$^{1}$ \and Xuanting Wu$^{1}$ \and Fei Ma$^{1}$\thanks{Corresponding author: \href{mailto:mafei@mail.buct.edu.cn}{mafei@mail.buct.edu.cn}.} \and Qiang Yin$^{1}$ \and Yuxin Hu$^{2}$\\[0.5em]
\small $^{1}$College of Information Science and Technology, Beijing University of Chemical Technology, Beijing 100029, China\\
\small $^{2}$Aerospace Information Research Institute, Chinese Academy of Sciences, Beijing 100190, China
}
\date{}

\begin{document}
\maketitle

\begin{abstract}
Synthetic aperture radar (SAR) image generation is challenging because target appearance varies strongly with observation geometry, whereas most generative methods operate primarily in the image domain and provide limited explicit geometric control. We propose PriSAR, a 3D geometric-prior-guided diffusion framework for parameter-controlled SAR image generation. PriSAR constructs viewpoint-aware point-cloud priors from 3D models through lightweight ray casting with a multi-bounce approximation and fuses them with textual and visual conditions through a Feature-wise Linear Modulation (FiLM)-based multi-modal fusion network. Low-rank adaptation (LoRA) efficiently adapts Stable Diffusion 3.5 Medium to the SAR domain. Experiments on a real SAR aircraft dataset show that the geometric prior improves structural fidelity and viewpoint consistency relative to representative generative baselines. We additionally report parameter sensitivity, computational cost, and data-augmentation experiments. These results indicate that PriSAR provides a practical geometry-guided approach for sparse-angle SAR data generation.
\end{abstract}


\vspace{0.8em}

\section{Introduction}

Synthetic Aperture Radar (SAR), as an active microwave imaging modality, supports all-time and all-weather Earth observation and has been widely used in applications such as reconnaissance, geological exploration, and disaster monitoring \citep{1, 2}. In recent years, deep learning (DL) has become an important paradigm for SAR image interpretation. Automatic Target Recognition (ATR) algorithms, particularly those based on Convolutional Neural Networks (CNNs) \citep{3} and Transformers \citep{4}, have shown strong performance. Unlike traditional approaches that rely on hand-crafted features, deep learning models learn hierarchical representations directly from data, improving the effectiveness of SAR processing \citep{5}. Beyond ATR, deep learning has also been applied to SAR image segmentation \citep{6, 7}, target detection \citep{8}, and multi-object tracking \citep{9}.

However, the effectiveness of these data-driven algorithms still depends strongly on the availability of sufficiently large and well-annotated datasets. In practical applications, SAR systems often face a pronounced data scarcity problem \citep{10}. On one hand, SAR imaging is highly sensitive to observation geometry (e.g., azimuth and depression angles), frequency bands, and polarization modes, so data collected under a specific operating condition are often limited. On the other hand, interpreting SAR imagery requires specialized domain expertise, which makes manual annotation expensive and time-consuming. As a result, deep models trained on limited SAR data are more susceptible to overfitting, especially in few-shot or zero-shot settings \citep{11}. In addition, SAR images are vulnerable to adversarial perturbations \citep{12}, which further increases the difficulty of building reliable recognition systems.

\begin{figure}[htbp]
    \centering
    \includegraphics[width=0.58\linewidth]{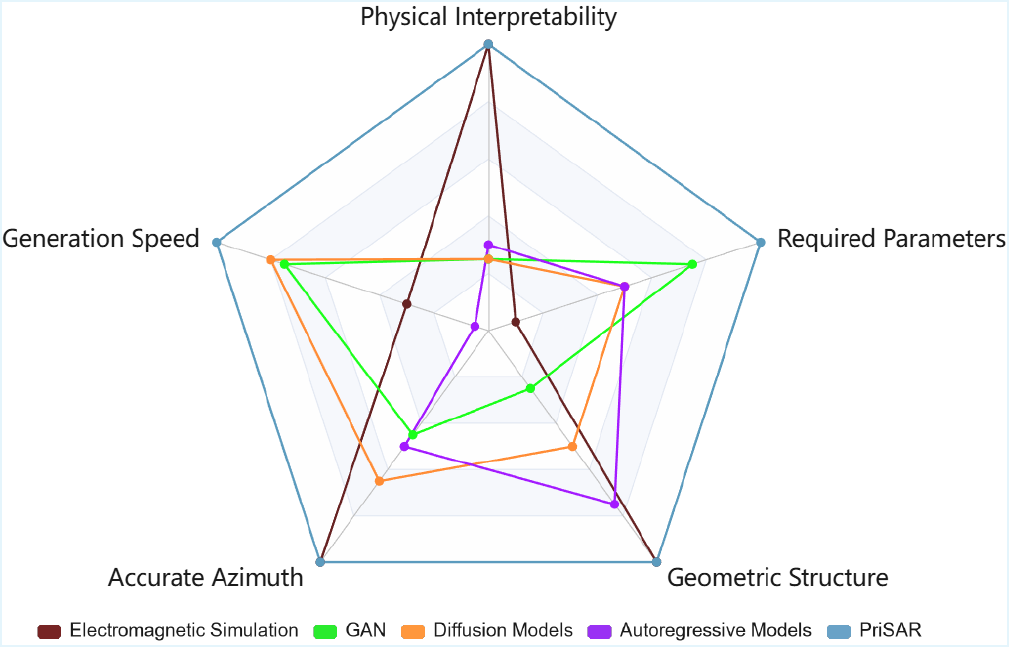}
    \caption{Conceptual radar-chart comparison of five representative SAR image generation paradigms. The axes denote generation speed, azimuth control, physical interpretability, parameter simplicity, and geometric structure preservation. The chart is intended to illustrate the relative positioning of different paradigms; PriSAR occupies an intermediate position between rigorous simulation and purely data-driven generation by combining geometric guidance with a diffusion backbone.}
    \label{fig:radarch}
\end{figure}

To mitigate this data shortage, researchers have explored a range of data augmentation strategies. As illustrated in the conceptual comparison in Figure \ref{fig:radarch}, these methods span a spectrum from physics-based simulation to data-driven generation, each with its own trade-off among efficiency, controllability, and interpretability.

\textbf{Electromagnetic Simulation.} Traditional electromagnetic simulation methods, notably Shooting and Bouncing Rays (SBR) and Physical Optics (PO), model radar echoes by solving Maxwell's equations or their approximations \citep{13,14,15}. GPU-accelerated SAR simulators have demonstrated that high geometric precision can be achieved under carefully specified physical settings \citep{15}. Although these methods offer strong physical interpretability and accurate geometric contours, their use in deep learning pipelines is limited by high computational cost and relatively limited textural diversity. They often produce overly smooth images that do not fully reproduce the speckle noise and background clutter in real SAR imagery, leading to a noticeable sim-to-real gap.

\textbf{Generative Adversarial Networks (GANs).} To reduce this gap, GANs \citep{17} became an important line of research. Frameworks such as Pix2Pix \citep{18} and CycleGAN \citep{19} have been used to translate optical images or semantic maps into the SAR domain. SAR-oriented GAN generators have also been explored for multi-view target synthesis, realism enhancement, and data augmentation \citep{16,20}. Despite their ability to synthesize plausible textures, GANs are known to suffer from unstable adversarial training and mode collapse \citep{21}, which can limit diversity. More importantly, conventional GANs do not explicitly model 3D observation geometry, so they may produce structurally implausible targets under changing viewpoints.

\textbf{Autoregressive Models.} Autoregressive (AR) models have also been explored for image synthesis. Methods such as VQ-VAE \citep{22} tokenize images into discrete codes and model the resulting sequence distribution, showing strong semantic modeling capability. DALL-E \citep{23} further demonstrated the potential of AR models in large-scale text-to-image generation. However, these models usually represent images as one-dimensional token sequences, which can increase inference cost and accumulate decoding errors. Their discrete latent formulation is also not naturally suited to the fine-grained geometric control required by SAR targets.

\begin{figure}[htbp]
    \centering
    \includegraphics[width=0.97\linewidth]{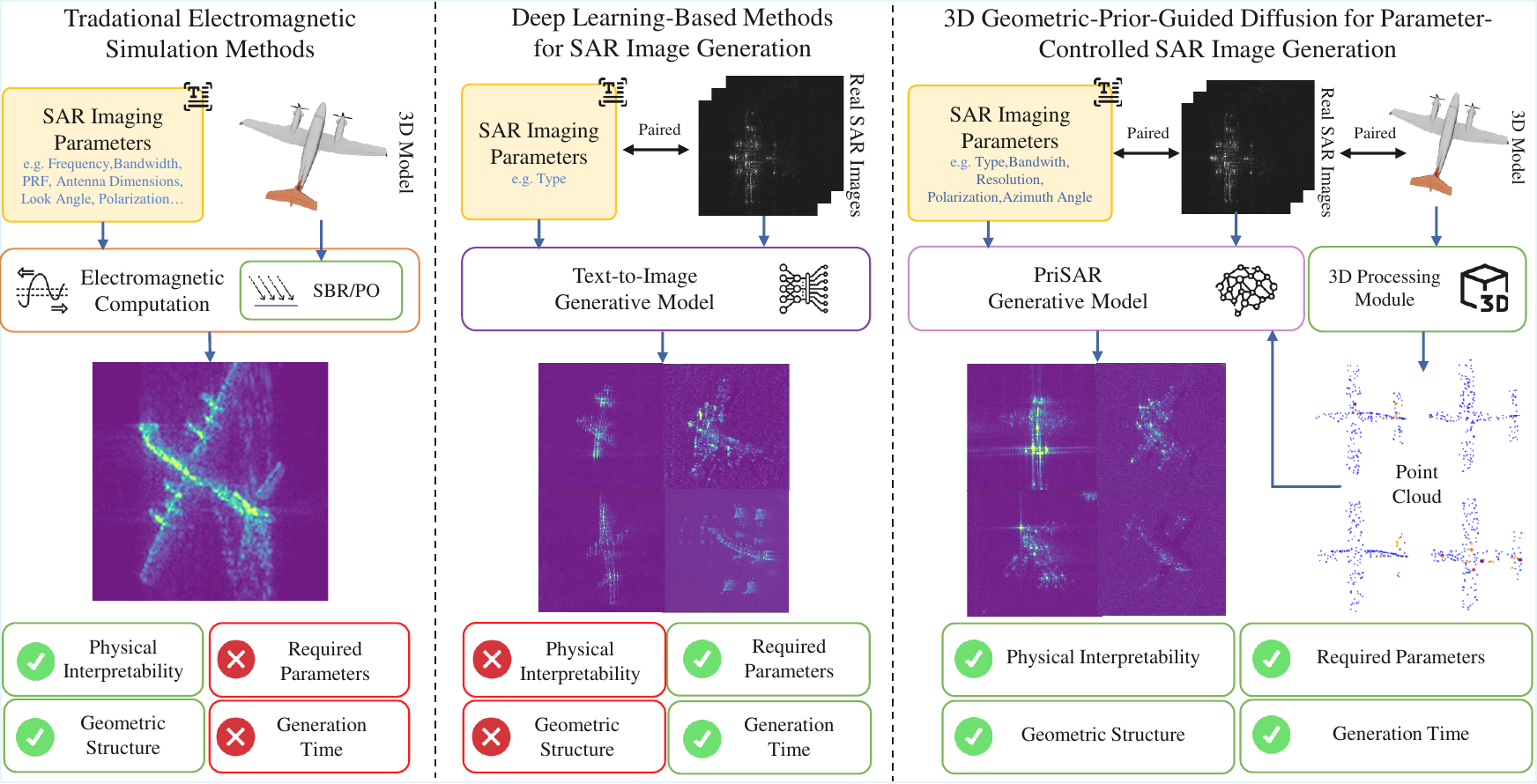}
    \caption{Illustration of the research motivation and paradigm comparison. \textbf{(Left)} Traditional electromagnetic simulations offer strong physical interpretability but are computationally expensive and require detailed parameter settings (e.g., antenna dimensions and PRF). \textbf{(Middle)} Generic text-to-image models provide efficient generation but often lack explicit geometric control, which may lead to structural inconsistency. \textbf{(Right)} PriSAR addresses this gap by extracting geometry-aware point-cloud priors from 3D models and using them to guide diffusion-based SAR generation.}
    \label{fig:mov}
\end{figure}

\textbf{Diffusion Models.} Recently, Denoising Diffusion Probabilistic Models (DDPM) \citep{24} and their variants (e.g., DDIM \citep{25}) have become a central family of generative models because of their strong distribution modeling ability. By learning a progressive denoising process, diffusion models can synthesize samples with high fidelity and diversity \citep{26}. In SAR and related conditional-generation settings, diffusion models have shown promising results in despeckling \citep{28}, remote sensing image generation \citep{31}, and general conditional image synthesis \citep{29}. However, directly applying text-to-image (T2I) models such as Stable Diffusion \citep{30} to SAR remains challenging. SAR-specific imaging effects, including layover and foreshortening, introduce a domain gap that natural-image models do not explicitly encode. As a result, text-only conditioning may produce targets with plausible radar-like textures but inconsistent geometric projections \citep{31}. Recent advances in polarimetric SAR systems \citep{32}, scattering-characteristic analysis \citep{33}, and physics-aware SAR generation \citep{51} further motivate the introduction of geometry-related priors into generative models. Figure \ref{fig:mov} summarizes this motivation.

While methods like ControlNet \citep{34} and T2I-Adapter \citep{35} introduce spatial conditions (Canny edges, depth maps), acquiring pixel-aligned semantic edge maps for SAR is impractical. Some pioneering works utilized 3D model projections \citep{36}, but naive 2D projections forfeit depth and scattering intensity information.

To address these challenges, we propose a 3D geometric-prior-guided diffusion framework for SAR image generation, termed \textbf{PriSAR}. The framework targets controllable SAR generation under sparse observation angles by integrating 3D-model-derived geometric priors with a diffusion backbone. Specifically, we use ray casting on 3D models to construct viewpoint-aware point clouds \citep{37}, which approximate scattering-center layouts and occlusion relationships under the given radar geometry. To fuse this multi-modal information, we introduce a multi-modal fusion network that jointly models geometric priors, image features, and textual descriptions \citep{38}. We further use Low-Rank Adaptation (LoRA) \citep{39} to fine-tune Stable Diffusion 3.5 Medium (SD3.5m) \citep{40} for the SAR domain in a parameter-efficient manner. Under the present setting, the resulting model produces samples whose azimuth-dependent structures more closely follow real SAR observations than text-only baselines, making it suitable for sparse-angle data generation and augmentation.

The main contributions are as follows.

\begin{itemize}
\item We propose PriSAR, a geometry-guided diffusion framework that introduces 3D-model-derived priors into SAR image generation. By explicitly incorporating viewpoint-aware geometric information, the framework supports controllability for observation-dependent SAR synthesis.

\item We design a viewpoint-aware geometric prior construction module and a geometry-text-image fusion strategy. The prior branch transforms 3D models into azimuth-aligned point-cloud representations through a lightweight, physically motivated projection process, while the FiLM-based fusion network dynamically combines these geometric cues with textual conditions and SAR visual features.

\item We conduct extensive experiments including visual-quality analysis, heuristic-parameter reporting, efficiency measurement, and application-oriented data augmentation experiments. The results indicate that geometric priors are beneficial for structural fidelity and viewpoint consistency, particularly under sparse-angle sampling conditions.
\end{itemize}

\section{Methodology}

\subsection{Overall Framework}

\begin{figure}[!t]
    \centering
    \includegraphics[width=1.0\textwidth]{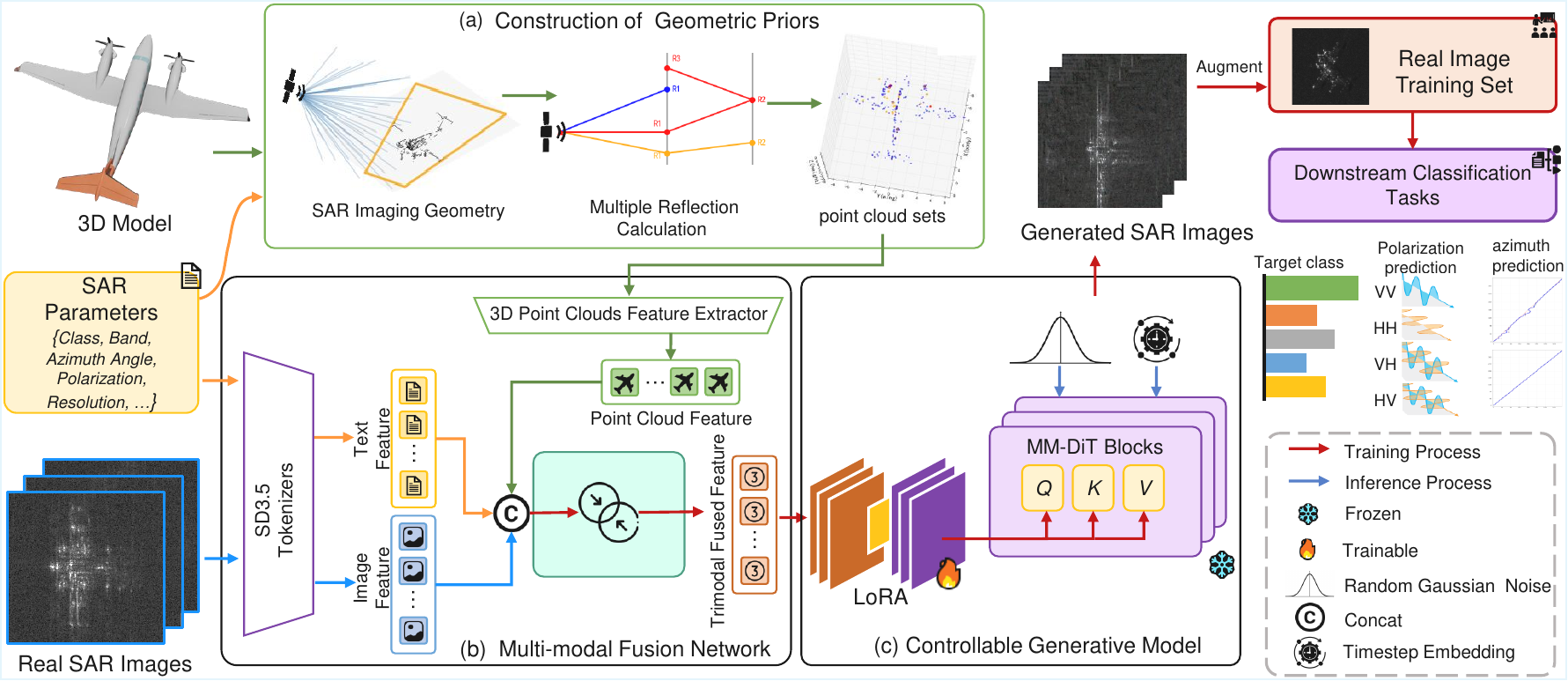} 
    \caption{Overall architecture of PriSAR. (a) \textbf{Construction of geometric priors}: 3D models are processed by a point-cloud module to extract view-dependent scattering cues, such as multi-bounce reflections and occlusion structures, which serve as geometric guidance. (b) \textbf{Multi-modal fusion network}: these features are combined with textual and visual conditions to guide the Stable Diffusion 3.5 Medium backbone, which is adapted to the SAR domain using LoRA while keeping most pre-trained weights frozen. (c) \textbf{Controllable generative model}: the generated SAR images can also be used in downstream classification experiments as an application-oriented evaluation.}
    \label{fig:overall}
\end{figure}

As illustrated in Figure \ref{fig:overall}, PriSAR combines a lightweight geometric-prior branch with latent diffusion generation. The geometric branch uses ray casting on 3D models under specified radar geometries (azimuth $\phi$, depression $\psi$) to approximate view-dependent scattering structures and occlusion relationships. These point clouds are encoded as geometry-aware representations and then fused with textual and visual conditions. We adopt LoRA to adapt the pre-trained SD3.5m backbone to the SAR domain in a parameter-efficient manner. Rather than claiming full electromagnetic reconstruction, the framework uses 3D-model-derived priors as controllable guidance that can enhance structural consistency during SAR generation.

\subsection{Construction of Geometric Priors}

\begin{figure}[htbp]
    \centering
    \includegraphics[width=0.29\linewidth]{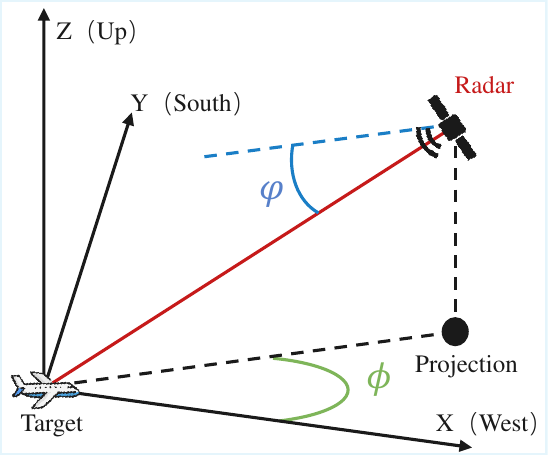} 
    \caption{Schematic diagram of object-centric radar observation geometry for point cloud simulation. The coordinate system sets the $X$-axis to West ($0^\circ$) and the $Y$-axis to South ($90^\circ$). The radar position is defined by azimuth $\phi$ and depression $\psi$, forming the LOS vector toward target center $O$. This setup supports constructing $\mathbf{T}_{radar}$ and performing ray casting for view-dependent scattering.}
    \label{fig:geo}
\end{figure}

To reduce the geometric inconsistency often observed in weakly constrained generative models, we construct an explicit geometric prior branch. The goal of this branch is not to synthesize SAR images directly, but to provide a geometry-aware approximation of view-dependent scattering structures and occlusion relationships. Because SAR is a side-looking ranging system, it produces characteristic geometric distortions such as foreshortening, layover, and shadow, which differ substantially from the perspective projection in natural optical images. To capture these effects, we use 3D models to perform virtual SAR scanning under an object-centric observation geometry, as illustrated in Figure \ref{fig:geo}.

In this geometric configuration, we define a specific coordinate system aligned with the target's orientation, where the $X$-axis is designated as West ($0^{\circ}$) and the $Y$-axis as South ($90^{\circ}$), with the azimuth angle increasing in a counter-clockwise direction. Based on this setup, for the 3D mesh model $\mathcal{M} = \{\mathcal{V}, \mathcal{F}\}$ of a given target (where $\mathcal{V}$ denotes the vertex set and $\mathcal{F}$ the face set), we construct the transformation matrix $\mathbf{T}_{radar}$ to map coordinates from the world system to the radar Line-of-Sight (LOS) coordinate system:

\begin{equation}
\mathbf{T}_{radar} = \begin{bmatrix} \cos\phi & -\sin\phi & 0 \\ \sin\phi\cos\psi & \cos\phi\cos\psi & -\sin\psi \\ \sin\phi\sin\psi & \cos\phi\sin\psi & \cos\psi \end{bmatrix}
\end{equation}

where $\phi$ represents the azimuth angle and $\psi$ denotes the depression angle.
To cover the target and simulate the SAR imaging region, we employ a two-level ray-generation strategy consisting of regular grid scanning and dense random scanning. Regular grid scanning generates the primary ray set with a fixed resolution $\Delta \theta$ in both azimuth and elevation to capture the overall contour of the target. For each azimuth step $\alpha$ and elevation step $\beta$, the ray direction vector $\vec{D}$ is given by:
\begin{equation}
\vec{D} = \begin{bmatrix} \cos(\beta) \cos(\alpha) \\ \cos(\beta) \sin(\alpha) \\ \sin(\beta) \end{bmatrix}
\end{equation}

Monte Carlo dense sampling is then used to fill the gaps left by regular sampling. To enhance detail capture for complex structures such as air intakes and landing gear, we introduce randomly perturbed rays directed toward the model center. For each ray $\mathbf{r}$, its direction vector $\mathbf{d}$ is defined as:

\begin{equation}
\mathbf{d} = \text{Normalize}(\mathbf{d}_{target} + \boldsymbol{\xi}), \quad \text{where } \boldsymbol{\xi} \sim \mathcal{N}(0, \sigma_{scan}^2 \mathbf{I})
\end{equation}
where $\mathbf{d}_{target}$ is the principal direction pointing to the geometric center of the model, $\boldsymbol{\xi}$ represents the Gaussian random perturbation term, and $\sigma_{scan}$ controls the beam divergence of the initial scan rays.

High-intensity regions in real SAR images often arise from secondary or tertiary reflections caused by corner-reflector-like structures. To approximate this effect, we design a recursive ray-tracing and energy-attenuation model. Recent work on scattering-characteristics analysis \citep{33} has shown the importance of multi-bounce responses for SAR target representation. The recursive tracing process models the dominant propagation paths before projecting the retained responses into the final point-cloud prior.

We define the ray state as $S_k = (\mathbf{p}_k, \mathbf{d}_k, E_k)$, representing the origin, direction, and energy intensity of the $k$-th bounce, respectively. For the $k$-th iteration, we first calculate the nearest intersection point $\mathbf{p}_{hit}$ between the ray and the mesh $\mathcal{M}$, as well as the surface normal vector $\mathbf{n}$ at that location. The pixel brightness in SAR images depends on the backscattering coefficient of the target. Based on the geometric properties at the intersection point, we construct a Pseudo-RCS Response Function $\Psi$ to update the propagated energy:
\begin{equation}
E_{k+1} = E_k \cdot e^{-\mu L_{path}} \cdot \Psi(\mathbf{p}_{hit}, \mathbf{n}, \mathcal{F})
\end{equation}
where $L_{path}$ denotes the cumulative path length, and $\mu$ represents the absorption coefficient. The response function $\Psi$ is composed of four geometric factors:
\begin{equation}
\Psi = \mathcal{W}_{base} \cdot \mathcal{H}_{edge}(\mathcal{F}) \cdot \mathcal{H}_{orient}(\mathbf{n}) \cdot \mathcal{H}_{struct}(\mathbf{p}_{hit})
\end{equation}
Here, $\mathcal{H}_{edge}$ is the edge-effect factor. When the area of the face containing the intersection point, $A_{\mathcal{F}}$, is smaller than the threshold $\tau_{area}$, typically corresponding to local structures such as wing edges or antennas, we set $\mathcal{H}_{edge} = \alpha_{edge} > 1.0$ to enhance the response. $\mathcal{H}_{orient}$ is an orientation factor that reflects the stronger response of side-looking radar to near-vertical surfaces. When the vertical component of the normal vector satisfies $|n_z| < \tau_{vert}$, a larger gain $\alpha_{vert}$ is assigned; otherwise a lower gain is used for more horizontal surfaces. $\mathcal{H}_{struct}$ is a structural factor that enhances regions such as wing-fuselage junctions to approximate dihedral corner-reflector effects through the gain $\alpha_{struct}$. The reflection direction of the secondary ray, $\mathbf{d}_{k+1}$, follows the law of specular reflection with an additional diffuse component caused by surface roughness:
\begin{equation}
\mathbf{d}_{spec} = \mathbf{d}_k - 2(\mathbf{d}_k \cdot \mathbf{n})\mathbf{n}
\end{equation}
\begin{equation}
\mathbf{d}_{k+1} = \text{Normalize}(\mathbf{d}_{spec} + \zeta \cdot \mathbf{u} + \boldsymbol{\eta}), \quad \mathbf{u} \sim \mathcal{U}(-1, 1)^3,\; \boldsymbol{\eta} \sim \mathcal{N}(0, \sigma_{reflect}^2 \mathbf{I})
\end{equation}
where $\zeta$ is the roughness coefficient and $\sigma_{reflect}$ controls the reflection-noise scale. This recursive process continues until the maximum number of reflections $K_{max}$ is reached or the propagated energy falls below a threshold.

Unless otherwise stated, the geometric prior branch uses the fixed heuristic parameters summarized in Table \ref{tab:param}. These values are manually selected to match the current acquisition setting and are then kept unchanged across aircraft categories, polarizations, and viewpoints in the reported experiments. In preliminary sensitivity checks, increasing $K_{max}$ from 1 to 3 and increasing $\zeta$ from 0 to 0.8 both improved SSIM and azimuth-related consistency, and the best setting was obtained at $K_{max}=3$ and $\zeta=0.8$, which is therefore used throughout this paper.

\begin{table}[t]
\footnotesize
\centering
\caption{Default heuristic parameters used in the geometric prior branch.}
\label{tab:param}
\begin{tabular}{
>{\centering\arraybackslash}p{1.4cm}
>{\centering\arraybackslash}p{1.0cm}
>{\raggedright\arraybackslash}p{3.6cm}
}
\toprule
Symbol & Value & Description \\
\midrule
$\alpha_{edge}$ & 2.0 & edge enhancement factor \\
$\alpha_{vert}$ & 2.0 & vertical-surface gain \\
$\alpha_{struct}$ & 1.6 & structural scattering gain \\
$\tau_{area}$ & 0.01 & small-face threshold \\
$\tau_{vert}$ & 0.3 & verticality threshold \\
$\zeta$ & 0.8 & diffuse scattering coefficient \\
$\mu$ & 0.03 & absorption coefficient \\
$K_{max}$ & 3 & maximum number of reflections \\
$\tau_{min}$ & 0.03 & reflection termination threshold \\
$\sigma_{scan}$ & 0.3 & initial ray perturbation scale \\
$\sigma_{reflect}$ & 0.1 & reflection noise scale \\
\bottomrule
\end{tabular}
\end{table}

\begin{figure}[!t]
    \centering
    \includegraphics[width=1.0\textwidth]{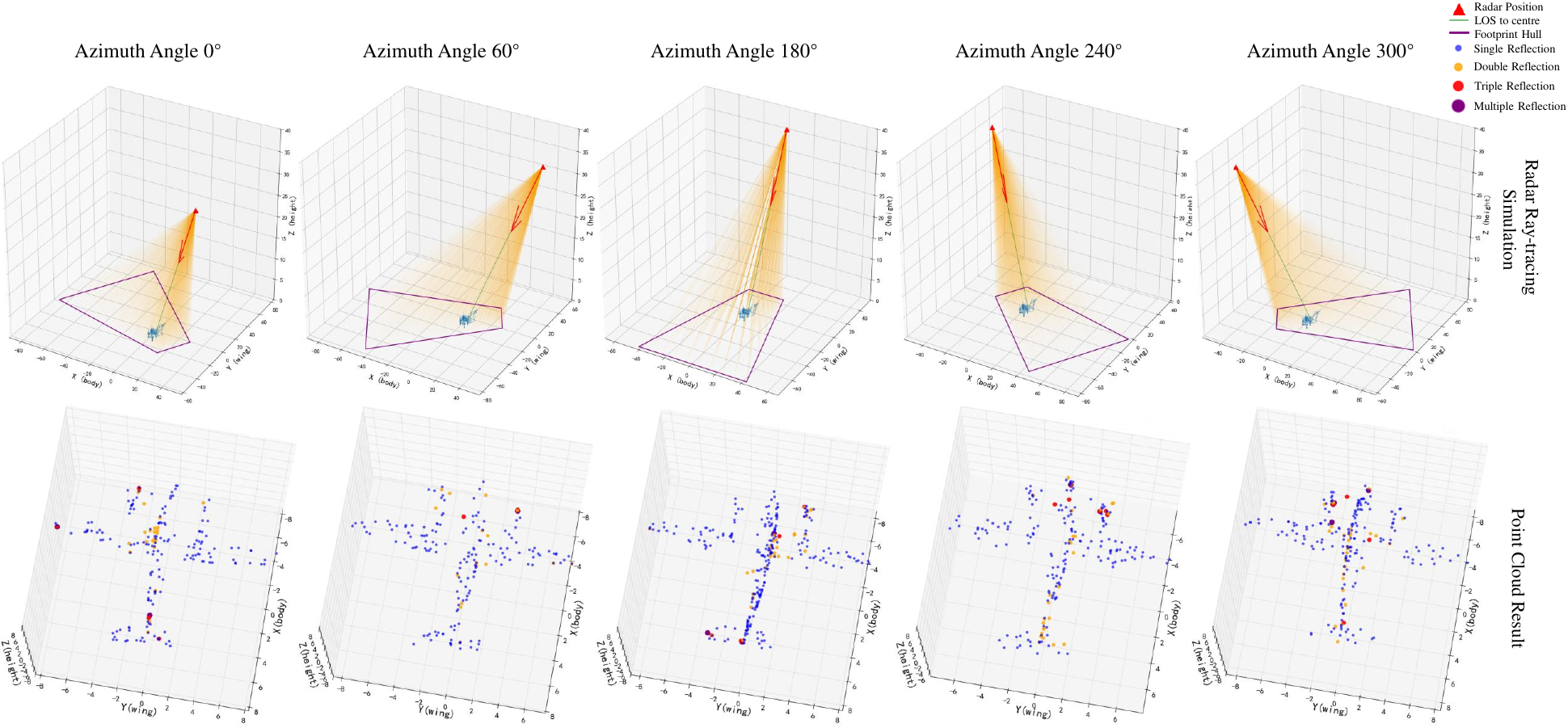} 
    \caption{Schematic diagram of radar illumination at different azimuth angles and the corresponding calculated point cloud models. The point clouds are color-coded by reflection type (blue for single, yellow for double, red for triple, and purple for multiple bounces), effectively capturing the view-dependent spatial distribution of strong scattering centers and occlusion relationships.}
    \label{fig:pc}
\end{figure}

As shown in Figure \ref{fig:pc}, the recursive procedure finally yields a point-cloud set $\mathcal{P} = \{(\mathbf{p}_{hit}, E_{final}, k) \mid E_{final} > \tau_{min}\}$ containing coordinates, energy, and reflection order. This point cloud describes the spatial distribution of single-, double-, and triple-bounce scattering responses and serves as a geometry-aware prior for the subsequent generative model.

To extract high-dimensional geometric features from the simulated SAR point cloud $\mathcal{P}$, we adopt Point Transformer V3 (PTv3) \citep{41} as the point-cloud encoder. Specifically, given the input point cloud $\mathcal{P} \in \mathbb{R}^{N \times (3+D_{in})}$ containing physical attributes such as intensity and reflection order, we first apply voxelization through Grid Sampling to reduce the impact of uneven local density. The PTv3 backbone then uses a serialization strategy based on space-filling curves to map unordered 3D points into an ordered 1D sequence. During feature extraction, the multi-scale attention blocks aggregate local geometric details and global shape context across different resolutions. The resulting feature $\mathbf{F}_{geo}$ encodes both the fine-grained 3D structure of the aircraft and the simulated scattering-intensity information. To adapt these features to the subsequent 2D generation network, we use adaptive projection mapping to project $\mathbf{F}_{geo}$ onto the radar slant-range plane, yielding a spatially aligned condition map $\mathbf{C}_{geo} \in \mathbb{R}^{H \times W \times C}$.

\subsection{Multi-modal Fusion Network}

\begin{figure}[!t]
    \centering
    \includegraphics[width=1.0\textwidth]{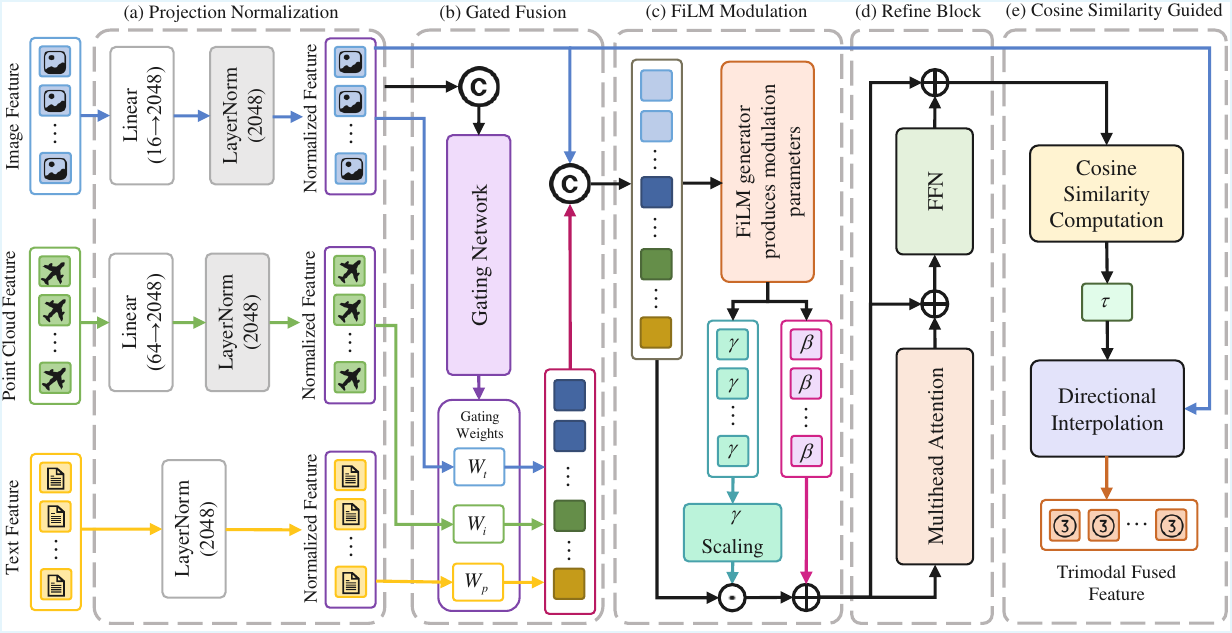} 
    \caption{The overall architecture of the proposed multi-modal fusion network. It comprises five cascaded modules: (1) \textbf{Projection Normalization} aligns multi-modal features into a unified high-dimensional space; (2) \textbf{Gated Fusion} dynamically assigns weights to text, image, and point cloud features based on information density; (3) \textbf{FiLM Modulation} injects fine-grained texture details through affine transformations; (4) \textbf{Refine Block} enhances feature interaction via multi-head attention; and (5) \textbf{Cosine Similarity Guided} mechanism enforces semantic consistency by interpolating features based on visual alignment.}
    \label{fig:fusion}
\end{figure}

To inject 3D geometric priors (point-cloud features) and visual conditions (image features) into the generative space of text-guided diffusion models, we design a multi-modal fusion network. Rather than using direct feature concatenation, the network adopts a five-stage cascaded architecture that progressively refines geometry consistency and semantic alignment, from coarse global weighting to fine-grained feature modulation. The overall workflow is illustrated in Figure \ref{fig:fusion}.

The text features $F_t \in \mathbb{R}^{B \times D}$ are extracted from the native SD3.5 text-conditioning branches, namely one T5 encoder and two CLIP encoders \citep{42,43}, with $D=2048$. In our implementation, the prompt follows a structured template such as ``SAR image, Pilatus PC 12, Ku-band, 0.5m resolution, VV polarization, azimuth 5 degree'', so that target category and imaging attributes are represented explicitly in the textual condition. The point-cloud features $F_p \in \mathbb{R}^{B \times 64}$ and image features $F_i \in \mathbb{R}^{B \times 16}$ lie in different feature spaces and have different dimensionalities. The image feature $F_i$ is extracted from the native SD image-conditioning pathway during training and acts as a visual anchor for SAR appearance alignment. We therefore first map the non-text modalities into a unified high-dimensional space by using linear projection together with Layer Normalization:
\begin{equation}
\tilde{F}_m = \text{LayerNorm}(\mathbf{W}_m F_m + \mathbf{b}_m), \quad m \in \{p, i\}
\end{equation}
\begin{equation}
\tilde{F}_t = \text{LayerNorm}(F_t)
\end{equation}
where $\tilde{F}_p, \tilde{F}_i$, and $\tilde{F}_t \in \mathbb{R}^{B \times 2048}$ denote the projected features after dimensional alignment.

To account for sample-dependent differences in the relative importance of physical priors and textual descriptions, we introduce an adaptive gating module that dynamically assigns modality weights. This step balances global complementarity across modalities. After concatenating the tri-modal features, the module predicts an importance score $\boldsymbol{\alpha}$ for each modality through a Multi-Layer Perceptron (MLP):
\begin{equation}
\boldsymbol{\alpha} = \text{Softmax}(\mathbf{W}_{gate} (\text{Dropout}(\phi(\mathbf{W}_{in}[\tilde{F}_t, \tilde{F}_p, \tilde{F}_i]))))
\end{equation}
where $\boldsymbol{\alpha} = [\alpha_t, \alpha_p, \alpha_i]$ and $\phi$ denotes the GELU activation function. Based on these weights, the preliminary fused feature $F_{pre}$ is computed as:
\begin{equation}
F_{pre} = \alpha_t \tilde{F}_t + \alpha_p \tilde{F}_p + \alpha_i \tilde{F}_i
\end{equation}

To inject image texture cues more finely into the generation process, we further introduce the Feature-wise Linear Modulation (FiLM) mechanism. Compared with gated additive fusion, FiLM modulates feature channels through affine transformation. We concatenate the preliminary fused feature $F_{pre}$ and the image feature $\tilde{F}_i$ to predict the scaling coefficients $\boldsymbol{\gamma}$ and shifting coefficients $\boldsymbol{\beta}$:
\begin{equation}
[\boldsymbol{\gamma}, \boldsymbol{\beta}] = \text{MLP}_{film}([F_{pre}, \tilde{F}_i])
\end{equation}
To improve training stability, we apply a Tanh constraint and residual scaling to the scaling coefficients:
\begin{equation}
\hat{\boldsymbol{\gamma}} = 1 + \lambda \cdot \tanh(\boldsymbol{\gamma})
\end{equation}
\begin{equation}
F_{film} = \hat{\boldsymbol{\gamma}} \odot F_{pre} + \boldsymbol{\beta}
\end{equation}
where $\odot$ denotes element-wise multiplication and $\lambda$ is a scaling factor set to 0.5. This mechanism allows the model to enhance or suppress specific feature channels according to image-texture cues.

To reduce semantic drift after multi-step feature transformation, we introduce a cosine-similarity-guided constraint. We calculate the cosine similarity $s_{cos}$ between the refined feature $F_{refined}$ and the image feature $\tilde{F}_i$. When the similarity falls below a preset threshold $\tau_{target}$, we interpolate the feature vector toward the image-semantic direction:
\begin{equation}
\tau = \text{Clamp}\left(\frac{\max(0, \tau_{target} - s_{cos})}{\max(\epsilon, 1 - s_{cos})}, 0, \tau_{max}\right)
\end{equation}
\begin{equation}
F_{final} = (1-\tau) \frac{F_{refined}}{\|F_{refined}\|_2} + \tau \frac{\tilde{F}_i}{\|\tilde{F}_i\|_2}
\end{equation}
where $\|\cdot\|_2$ denotes the $\ell_2$ norm. Finally, feature energy is restored by amplitude recovery. This strategy encourages the generated SAR images to remain consistent with the input geometry and visual conditions while preserving textual semantics.

\subsection{Controllable Generative Model}

\begin{figure}[htbp]
    \centering
    \includegraphics[width=0.38\linewidth]{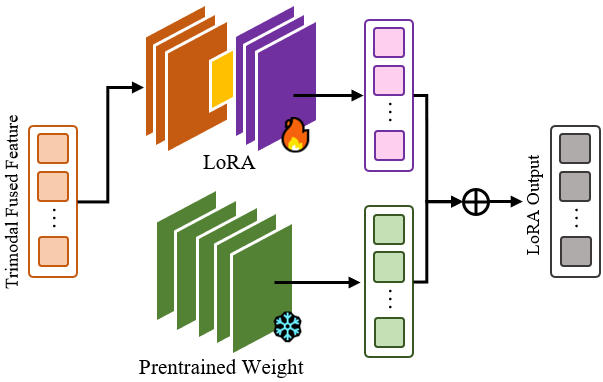} 
    \caption{The schematic diagram of the LoRA working principle applied in the SD3.5 backbone.}
    \label{fig:lora}
\end{figure}

Considering the large parameter count of SD3.5m and the scarcity of labeled SAR data, full fine-tuning would be computationally expensive and more prone to overfitting. We therefore adopt a parameter-efficient fine-tuning strategy. During training, we freeze the pre-trained VAE encoder $\mathcal{E}$, decoder $\mathcal{D}$, and the CLIP/T5 text encoders to preserve the original latent space and semantic representations. The trainable parameter set $\Theta$ contains only two components:
\begin{equation}
\Theta = \{\theta_{LoRA}, \theta_{Fusion}\}
\end{equation}
Here, $\theta_{LoRA}$ denotes the Low-Rank Adaptation matrices inserted into the attention layers of the MMDiT backbone to adapt the model to SAR-domain texture statistics, while $\theta_{Fusion}$ denotes the parameters of the proposed multi-modal fusion network that learns nonlinear interactions among text, point-cloud, and image conditions. The schematic principle of LoRA is shown in Figure \ref{fig:lora}.

The training objective follows the standard noise-prediction formulation of latent diffusion models. Given a real SAR image $x \in \mathcal{D}_{train}$, we first use the VAE encoder to map it into the latent space, $z_0 = \mathcal{E}(x)$. Gaussian noise $\epsilon \sim \mathcal{N}(0, \mathbf{I})$ is then added to $z_0$ at time step $t \sim \mathcal{U}(\{1, \dots, T\})$ to obtain the noisy latent variable $z_t$. The network $\epsilon_\theta$ takes as input the noisy latent $z_t$, the time step $t$, and the multi-modal condition embedding $C_{fused}$ produced by the fusion network. The optimization objective minimizes the Mean Squared Error (MSE) between the predicted noise and the sampled noise:
\begin{equation}
\mathcal{L}_{simple} = \mathbb{E}_{z_0, \epsilon, t, C_{fused}} \left[ \left\| \epsilon - \epsilon_\theta(z_t, t, C_{fused}) \right\|_2^2 \right]
\end{equation}
where $C_{fused}$ is the fused condition derived from text features $F_t$, geometric features $F_p$ generated by SAR simulation, and the corresponding SAR image features $F_i$. Through end-to-end optimization, $\theta_{Fusion}$ learns how to adjust the contribution of the geometric prior, while $\theta_{LoRA}$ adapts the backbone to SAR-specific texture and clutter statistics.

\begin{figure}[htpb]
    \centering
    \includegraphics[width=0.97\textwidth]{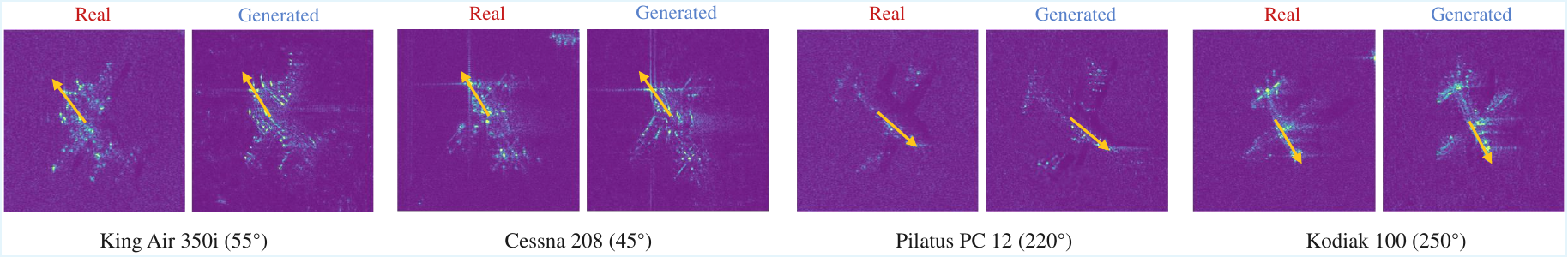} 
    \caption{Qualitative generation results of PriSAR across different aircraft categories and azimuth angles. The figure presents side-by-side comparisons of real SAR images and generated samples for King Air 350i ($55^{\circ}$), Cessna 208 ($45^{\circ}$), Pilatus PC 12 ($220^{\circ}$), and Kodiak 100 ($250^{\circ}$). Guided by 3D geometric priors, the generated targets remain consistent with the specified viewpoints and reduce the azimuth ambiguity often observed in purely data-driven approaches.}
    \label{fig:azgen}
\end{figure}

During inference, PriSAR operates in a ``Text + Geometry'' conditioning mode and does not require paired image inputs. Users provide a text instruction $T_{in}$ containing the target category and imaging parameters. The system retrieves the corresponding 3D model according to the category specified in $T_{in}$, parses the azimuth and depression angles, invokes the SAR simulator described in Section 2.2 to generate the point cloud $P$ under the target viewpoint, and extracts the geometric feature $F_p$. Because no paired SAR image is available during inference, the auxiliary image branch used in training is removed at test time, and generation is driven mainly by text and geometric conditions. In this sense, the image feature branch functions as a training-time visual anchor that regularizes texture-semantic alignment rather than as a required modality during inference, so its removal changes only an auxiliary cue instead of the principal conditioning pathway. To balance fidelity and diversity, we further employ Classifier-Free Guidance (CFG). During sampling, we compute both the conditional predicted noise $\epsilon_\theta(z_t, t, C_{fused})$ and the unconditional predicted noise $\epsilon_\theta(z_t, t, \emptyset)$, where the latter uses empty text and empty point-cloud inputs. The final denoising step is given by:
\begin{equation}
\tilde{\epsilon}_t = \epsilon_\theta(z_t, t, \emptyset) + w \cdot \left(\epsilon_\theta(z_t, t, C_{fused}) - \epsilon_\theta(z_t, t, \emptyset)\right)
\end{equation}

where $w$ is the guidance scale. A larger $w$ encourages the model to follow the 3D point-cloud contours more closely, whereas a smaller $w$ allows more textural variation. Finally, the denoised latent variable $z_0$ is mapped back to pixel space through the VAE decoder $\mathcal{D}$ to obtain the final SAR image. As shown in Figure \ref{fig:azgen}, PriSAR generates images whose azimuth-dependent structures are more consistent with those of real SAR samples.

\section{Experiments}

\subsection{Experimental Setup}

To characterize the performance of PriSAR, this section evaluates the method from two complementary perspectives. First, we use visual-quality metrics to quantify the fidelity and diversity of the synthesized images. Second, we conduct downstream classification experiments to assess the application value of the generated samples for data augmentation.

\begin{figure}[b!]
    \centering
    \includegraphics[width=0.99\linewidth]{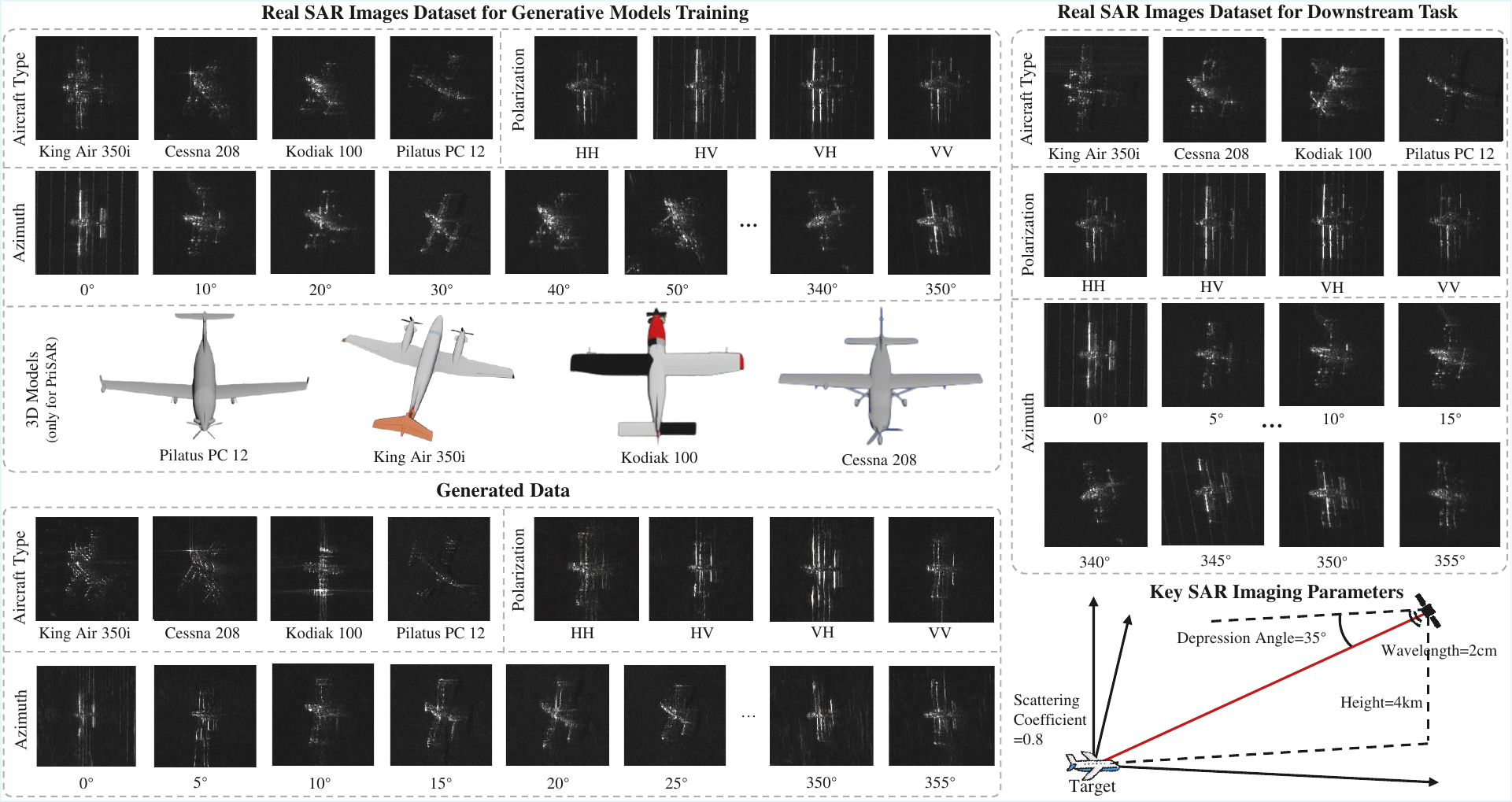} 
    \caption{Data structure and composition strategy. The figure illustrates the three key components of our experimental data: 
    (1) \textbf{Real SAR Images Dataset for Generative Models Training}: A sparse subset of real data sampled at coarse $10^{\circ}$ azimuth intervals (e.g., $0^{\circ}, 10^{\circ}$), used to supervise the generative baseline models.
    (2) \textbf{Generated Data}: Synthetic samples produced by PriSAR (guided by \textbf{3D Models}) specifically at the missing intermediate $5^{\circ}$ azimuths (e.g., $5^{\circ}, 15^{\circ}$), designed to fill the angular gaps.
    (3) \textbf{Real SAR Images Dataset for Downstream Task}: A reserved split of the real SAR dataset (saved from the initial partitioning) which is subsequently merged with the generated data to construct the final augmented training set for the downstream classification task.}
    \label{fig:data}
\end{figure}

\subsubsection{Dataset}
To evaluate geometry-guided SAR generation, we constructed a high-resolution SAR aircraft dataset from field collection campaigns at Shanxi Yaocheng Airport. The full dataset contains 8,536 images from four aircraft categories, namely Cessna 208, Kodiak 100, King Air 350i, and Pilatus PC 12, as well as four polarization modes: HH, HV, VH, and VV. For generative-model training, we use a sparse subset of 1,870 images sampled at $10^{\circ}$ azimuth intervals, while the remaining 6,666 images are reserved for evaluation on a denser $5^{\circ}$ azimuth grid. Because polarimetric SAR systems \citep{32} and sub-look decomposition techniques \citep{8} can enhance target discrimination, we report results across multiple polarization modes.

As illustrated in Figure \ref{fig:data}, the real data are partitioned into two groups to simulate a sparse-angle setting. For generative-model training, we use only the sparse subset sampled at $10^{\circ}$ azimuth intervals (e.g., $0^{\circ}, 10^{\circ}, 20^{\circ}, \dots$). PriSAR is then used to generate synthetic samples at the missing intermediate $5^{\circ}$ azimuths (e.g., $5^{\circ}, 15^{\circ}, 25^{\circ}, \dots$). These generated images are combined with the sparse real data to form a denser dataset for downstream classification experiments.

\subsubsection{Implementation Details}
Our framework was implemented in PyTorch and trained on a cluster equipped with three NVIDIA GeForce RTX 4090 GPUs. We initialized the generative backbone with SD3.5m weights. To satisfy computational constraints and improve training stability, all SAR images were resized to $256 \times 256$ pixels. We adopted LoRA fine-tuning with rank 32 and alpha 16, while keeping the VAE and pre-trained text encoders frozen. Training was performed in BF16 mixed precision to reduce memory usage. We used the Prodigy optimizer with a D-coefficient of 0.5 and an initial learning rate $d_0$ of $1 \times 10^{-4}$. The trainable LoRA and fusion parameters were optimized with cosine annealing, and the learning rate of the denoising backbone branch was set to $1 \times 10^{-3}$. The total batch size was 12.

\subsubsection{Evaluation Metrics}
To assess the quality of the generated SAR imagery, we report four complementary metrics spanning pixel-level fidelity and perceptual similarity. Peak Signal-to-Noise Ratio (PSNR) and Structural Similarity Index Measure (SSIM) evaluate low-level reconstruction quality and structural preservation, respectively. We further compute the Learned Perceptual Image Patch Similarity (LPIPS) and the Fr\'echet Inception Distance (FID) as auxiliary perceptual indicators for generative-model comparison. Following recent discussions on SAR generative evaluation \citep{49,50}, we do not interpret LPIPS or FID in isolation, because their feature extractors are primarily trained on natural-image corpora. Instead, we analyze them jointly with PSNR, SSIM, viewpoint consistency, and downstream utility.

\subsubsection{Downstream Classification Model}
We use the PyTorch Image Models Multi-Label Classification framework \citep{44} as the downstream testbed. Built on the timm ecosystem, this framework provides a unified interface for deep classification backbones, including convolutional neural networks and vision transformers. Using this codebase helps maintain a consistent training and evaluation protocol across different augmentation settings. In this experiment, we examine whether the synthetic samples generated by PriSAR can serve as useful additional training data for classifier development on real SAR datasets. Specifically, the generated images are mixed with limited real training samples, and the classifiers are then evaluated on a held-out set of real SAR images.

\subsection{Efficiency Analysis}

Besides image quality, we also measure the computational overhead introduced by the geometric prior branch. Table \ref{tab:efficiency} reports the average running time over 100 repeated trials or images. Constructing one point-cloud prior takes 48.3 seconds per azimuth on CPU. The subsequent PriSAR inference takes 13.1 seconds per image on an NVIDIA RTX 4090 with 30 sampling steps and approximately 10 GB GPU memory. Under the same setting, the SD3.5m baseline without geometric priors requires 11.3 seconds per image. Therefore, the main additional cost of PriSAR lies in prior construction, whereas the denoising stage remains close to the backbone model. This trade-off is computationally more tractable than rigorous electromagnetic simulation while still providing geometry-aware guidance beyond text-only generation.

\begin{table}[htbp]
\footnotesize
\caption{Average running cost measured over 100 runs.}
\label{tab:efficiency}
\centering
\begin{tabular}{ll}
\toprule
Item & Measurement \\
\midrule
Point-cloud prior construction & 48.3 s per azimuth on CPU \\
PriSAR inference & 13.1 s per image on RTX 4090 \\
SD3.5m inference & 11.3 s per image on RTX 4090 \\
\bottomrule
\end{tabular}
\end{table}

\begin{figure}[htbp]
    \centering
    \includegraphics[width=0.96\textwidth]{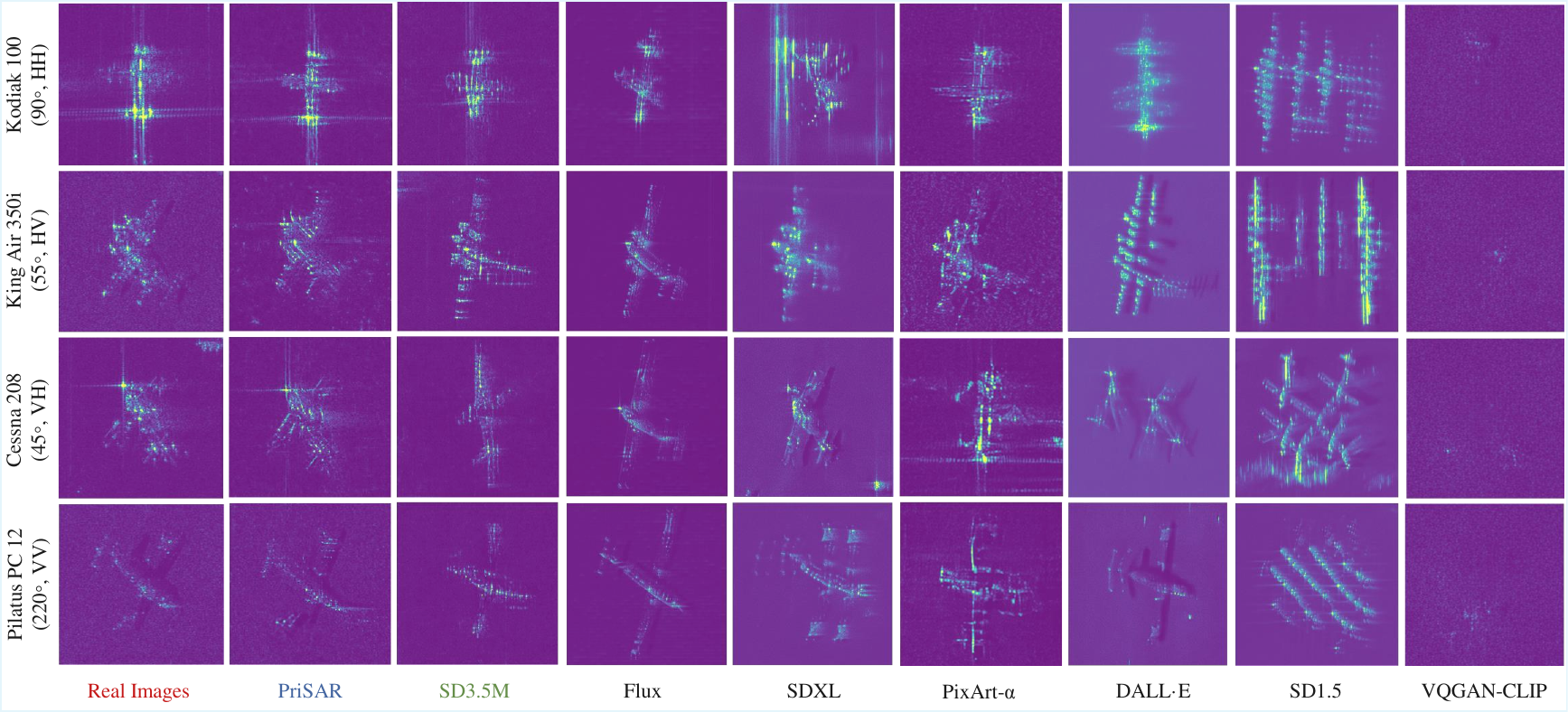} 
    \caption{Visual quality comparison between PriSAR and representative baseline models. Each column corresponds to a distinct configuration: column 1 shows Kodiak 100 ($90^{\circ}$, HH); column 2 shows King Air 350i ($55^{\circ}$, HV); column 3 shows Cessna 208 ($45^{\circ}$, VH); and column 4 shows Pilatus PC 12 ($220^{\circ}$, VV). The generated images generally follow the designated observation viewpoints more closely and preserve more stable target structures than the baselines.}
    \label{fig:cmp}
\end{figure}

\subsection{Visual Quality Evaluation}

As shown in Figure \ref{fig:cmp}, we compare PriSAR with representative baselines covering several major generative paradigms. These include GAN-based methods represented by VQGAN-CLIP \citep{45,42}, autoregressive models represented by DALL-E \citep{23}, and several recent diffusion models, including SD1.5 \citep{30}, SDXL \citep{47}, PixArt \citep{46}, Flux \citep{48}, and SD3.5m \citep{40}. Here the "SD3.5m" entry denotes the vanilla Stable Diffusion 3.5 Medium backbone fine-tuned on the SAR dataset without 3D point-cloud features. This setting serves as a direct ablation baseline for quantifying the contribution of the geometric prior branch.

\begin{figure}[htbp]
    \centering
    \includegraphics[width=0.95\textwidth]{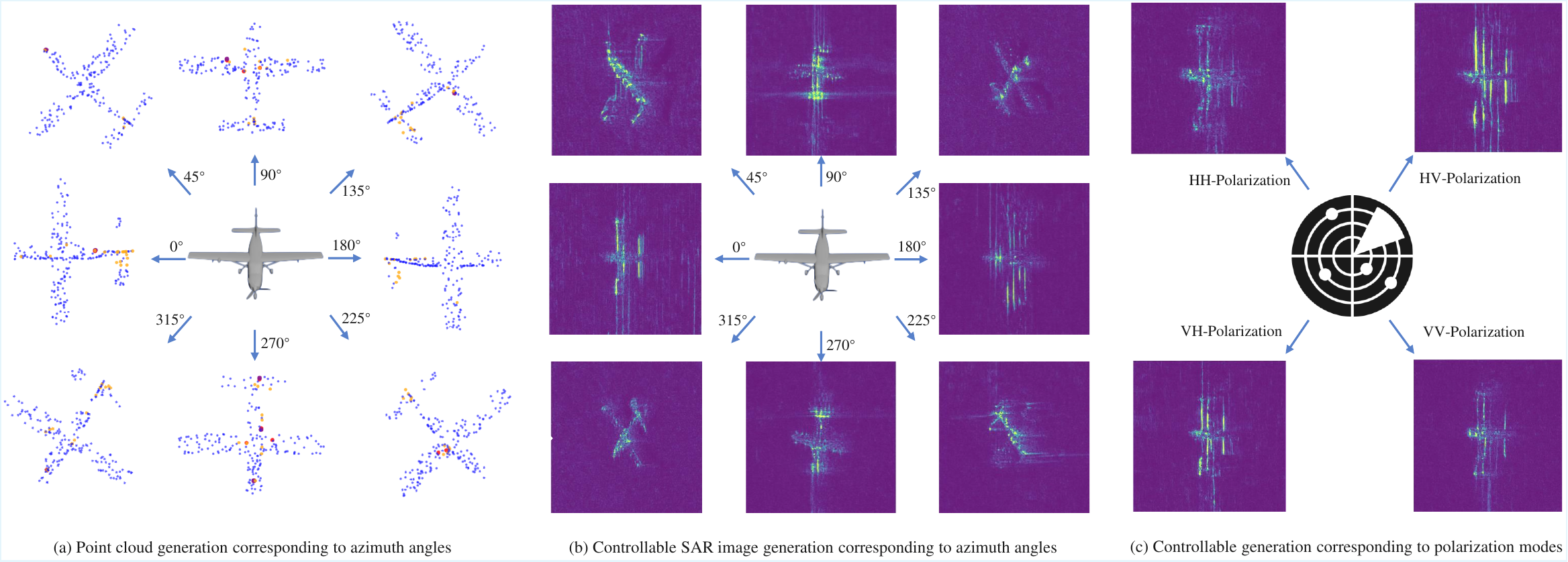} 
    \caption{Visualization of the multi-modal correspondence between text, point clouds, and generated images. (a) \textbf{Point cloud generation corresponding to azimuth angles}: The intermediate 3D point cloud priors generated under varying observation geometries (e.g., $0^{\circ}$ to $315^{\circ}$), serving as explicit geometric guidance. (b) \textbf{Controllable SAR image generation corresponding to azimuth angles}: The synthesized SAR images that follow the specific azimuths defined in (a), demonstrating clear geometric alignment with the provided prompts. (c) \textbf{Controllable generation corresponding to polarization modes}: The generation results under different polarization conditions (e.g., HH, HV, VH, VV) reflect polarization-dependent scattering differences in the synthesized SAR images.}
    \label{fig:tpi}
\end{figure}

To further illustrate the controllable generation process, we visualize the correspondence among the input \textbf{Textual Prompts} (specifying attributes such as azimuth angle and polarization mode), the intermediate \textbf{3D Point Cloud Priors} (providing geometric guidance), and the final \textbf{Generated SAR Images} in Figure \ref{fig:tpi}. The generated images generally remain consistent with the point-cloud-defined structures while also reflecting polarization-dependent appearance changes, which is consistent with the intended role of the geometry-guided mechanism.

\subsubsection{Evaluation Protocol}
To compare visual quality under consistent conditions, we establish a standardized label-to-image evaluation protocol. All competing models are trained or fine-tuned on the same SAR aircraft data partition. During evaluation, semantic labels from the held-out test set are used as input conditions. For each real SAR image in the test set, the trained model generates a corresponding synthetic sample from the same textual description. This paired evaluation protocol allows us to assess how closely each model reproduces SAR textures and geometry under matched semantic conditions.

\subsubsection{Evaluation Results}
Table \ref{tab1} presents the quantitative comparison results on the real SAR dataset. PriSAR achieves the highest PSNR and SSIM under the present protocol, suggesting comparatively stronger reconstruction fidelity and structural preservation. Relative to the vanilla SD3.5m baseline, the proposed geometric prior also yields lower LPIPS and FID, which is consistent with better distributional alignment. Considering the domain gap between natural-image feature spaces and SAR imagery, we regard LPIPS and FID here as supportive evidence rather than standalone proof of SAR realism.

\begin{table}[htbp]
\footnotesize
\caption{Quantitative comparison of visual quality metrics for different generative models on the real SAR dataset. The best results are in bold, and the second-best results are underlined.}
\label{tab1}
\centering
\begin{tabularx}{\textwidth}{YYYYY} 
\toprule
  Methods & PSNR $\uparrow$ & SSIM $\uparrow$ & LPIPS $\downarrow$ & FID $\downarrow$ \\
  \midrule
  VQGAN-CLIP & 8.79 & 0.188 & 0.561 & 35.2 \\
  SD1.5      & 11.15 & 0.378 & 0.472 & 18.7 \\
  DALL-E     & 23.13 & 0.505 & 0.310 & 6.1 \\
  PixArt     & 20.36 & 0.428 & 0.278 & 7.5 \\
  SDXL       & 15.04 & 0.592 & 0.352 & 10.8 \\
  Flux       & \underline{26.18} & 0.715 & \underline{0.251} & \underline{4.1} \\
 SD3.5m     & 25.23 & \underline{0.738} & 0.265 & 5.5 \\
\textbf{PriSAR} & \textbf{31.36} & \textbf{0.812} & \textbf{0.232} & \textbf{3.4} \\
\bottomrule
\end{tabularx}
\end{table}

\subsection{Downstream Classification Evaluation}

\begin{table}[!t]
\footnotesize
\caption{Performance comparison on downstream multi-label classification tasks. The best results are in bold, and the second-best results are underlined.}
\label{tab2}
\setlength{\tabcolsep}{3pt}
\renewcommand{\arraystretch}{0.95}
\centering
\begin{tabularx}{\textwidth}{cYYYYYYYYY}
\toprule
\multirow{2}{*}{Methods} & \multicolumn{3}{c}{Aircraft Type} & \multicolumn{3}{c}{Azimuth} & \multicolumn{3}{c}{Polarization} \\ 
\cmidrule(lr){2-4} \cmidrule(lr){5-7} \cmidrule(lr){8-10} 
 & Precision & Recall & F1-Score & Precision & Recall & F1-Score & Precision & Recall & F1-Score \\ \midrule
Real Image & 0.883 & 0.903 & 0.887 & 0.078 & 0.225 & 0.109 & 0.664 & 0.668 & 0.665 \\       
VQGAN-CLIP & 0.650 & 0.623 & 0.621 & 0.038 & 0.120 & 0.050 & 0.429 & 0.438 & 0.425 \\
SD1.5 & 0.824 & 0.867 & 0.823 & 0.056 & 0.177 & 0.077 & 0.661 & 0.659 & 0.628 \\
DALL-E &\underline{0.998} & \underline{0.998} & \underline{0.998} & 0.671 & 0.657 & 0.641 & 0.765 & 0.769 & 0.766 \\
PixArt & 0.997 & 0.995 & 0.996 & 0.491 & 0.469 & 0.437 & 0.743 & 0.745 & 0.729 \\
SDXL & 0.992 & 0.986 & 0.989 & 0.373 & 0.390 & 0.327 & 0.745 & 0.738 & 0.715 \\
Flux & 0.996& 0.994 & 0.995 & 0.786 & 0.773 & 0.769 & \underline{0.766} & \underline{0.765} & \underline{0.763} \\
SD3.5m & 0.994 & 0.994 & 0.994 & \underline{0.797} & \underline{0.787} & \underline{0.782} & 0.735 & 0.735 & 0.731 \\ 
\textbf{PriSAR} & \textbf{1.000} & \textbf{1.000} & \textbf{1.000} & \textbf{0.940} & \textbf{0.940} & \textbf{0.939} & \textbf{0.933} & \textbf{0.934} & \textbf{0.933} \\ 
\bottomrule
\end{tabularx}
\renewcommand{\arraystretch}{1.0}
\end{table}

\subsubsection{Experimental Pipeline and Data Partitioning}
\begin{figure}[htbp]
    \centering
    \includegraphics[width=0.97\linewidth]{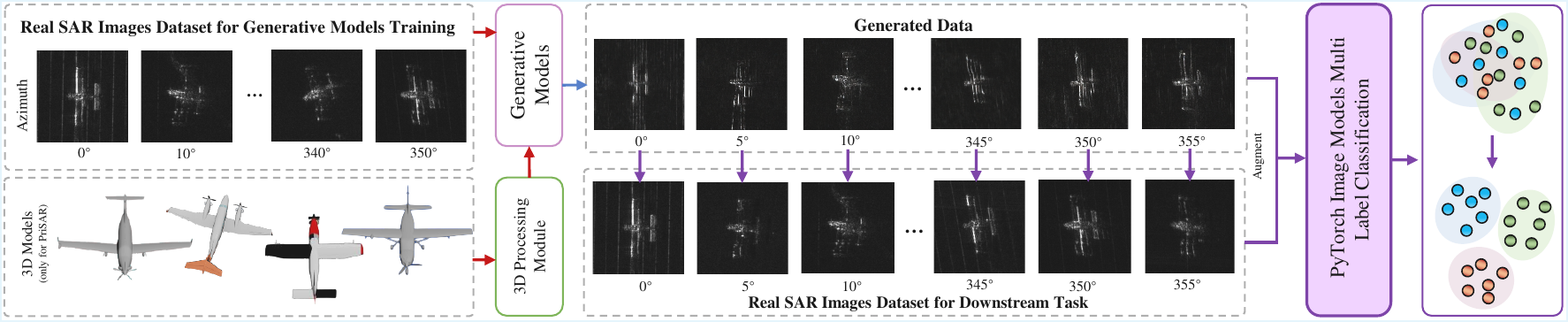} 
    \caption{Workflow of the downstream evaluation. Sparse $10^{\circ}$ real data are first used to train the generative models; PriSAR then synthesizes the missing intermediate azimuths with 3D-guided priors, and the resulting mixed dataset is used for downstream classification.}
    \label{fig:downstream}
\end{figure}

To evaluate the utility of PriSAR for dense angular completion under data scarcity, we design the downstream classification experiment shown in Figure \ref{fig:downstream}. We simulate a sparse-data scenario by downsampling the available real training data to coarse $10^{\circ}$ azimuth intervals (e.g., $0^{\circ}, 10^{\circ}, 20^{\circ}, \dots$). This sparse subset serves as the sole training supervision for the generative models.

During data augmentation, PriSAR generates synthetic SAR images at the missing intermediate $5^{\circ}$ azimuth offsets (e.g., $5^{\circ}, 15^{\circ}, 25^{\circ}, \dots$). These geometry-guided synthetic samples are then interleaved with the sparse real data to reconstruct a denser dataset with a $5^{\circ}$ angular resolution. The resulting mixed dataset is used to train the \textit{PyTorch Image Models Multi-Label Classifier}, and its performance is evaluated on a held-out test set to examine the effect of fine-grained angular completion.

While visual fidelity metrics quantify image quality, they do not necessarily reflect semantic correctness or discriminative utility. We therefore include a downstream classification task as an application-oriented evaluation. Using the same PyTorch Image Models framework, we compare different augmentation settings under the current sparse-angle scenario. As shown in Table \ref{tab2}, PriSAR achieves the best overall downstream results among the compared generative models under the present setting.

\begin{figure}[t!]
    \centering
    \includegraphics[width=0.95\linewidth]{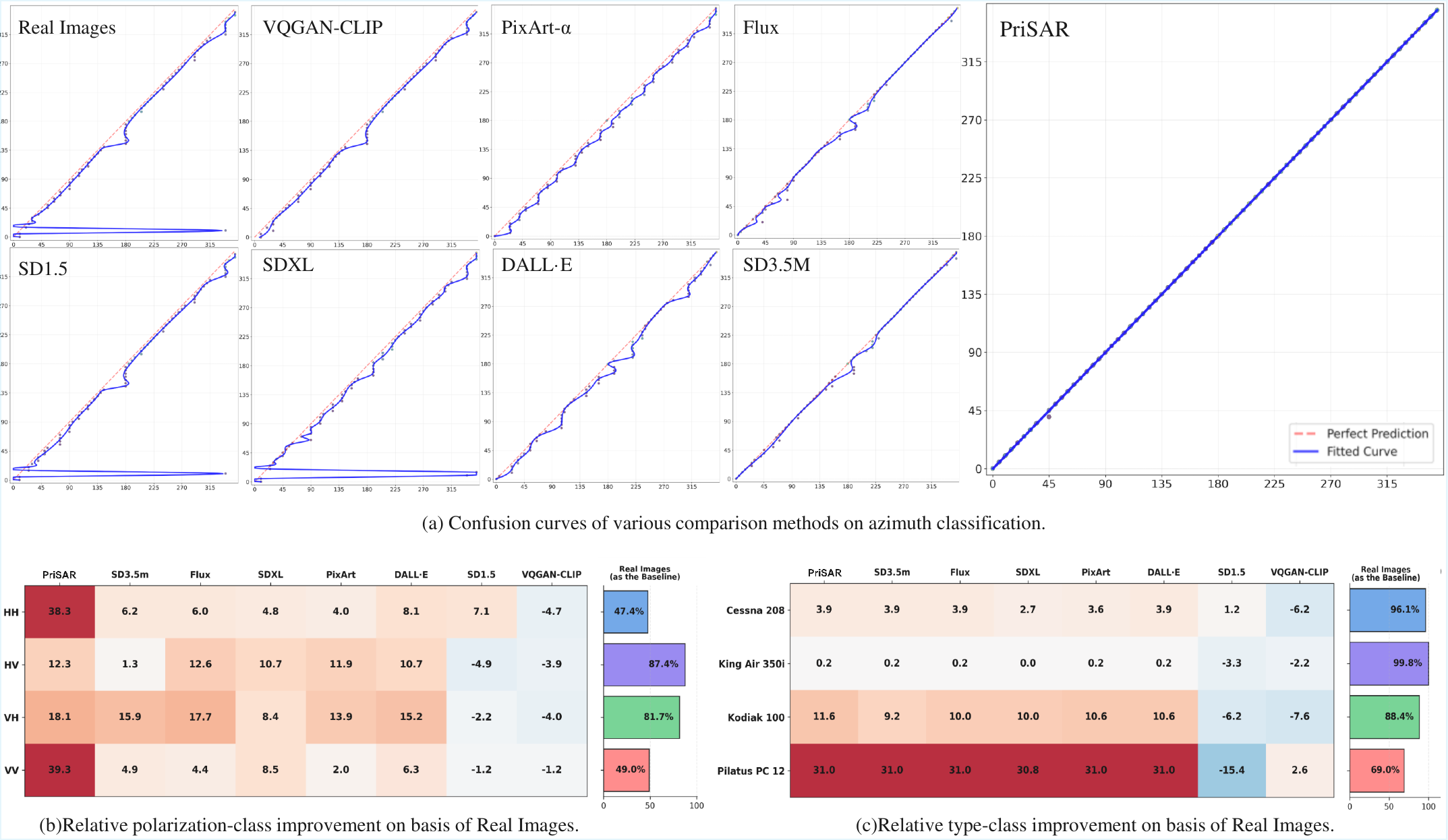} 
    \caption{Visualization of downstream classification results: (a) azimuth confusion curves, (b) relative polarization-class gains over the real-data baseline, and (c) relative aircraft-type gains. PriSAR appears to preserve more discriminative cues than the compared baselines.}
    \label{fig:ex}
\end{figure}

To provide a more detailed analysis, we visualize the downstream classification results in Figure \ref{fig:ex}. The azimuth confusion curves in Figure \ref{fig:ex}(a) show that baseline methods such as SD3.5m and Flux still exhibit noticeable angular ambiguity, whereas PriSAR follows the ground-truth trend more smoothly. The heatmaps in the lower panels further show relative gains across several aircraft categories and polarization modes, suggesting that geometry-guided generation preserves useful semantic cues for downstream recognition.

\subsubsection{Aircraft Type Classification}
In the coarse-grained aircraft type classification task, recent models such as Flux and SD3.5m already achieve F1-scores above 0.99, while PriSAR reaches 1.000. This result suggests that the generated samples preserve the main morphological distinctions among aircraft categories.

\subsubsection{Polarization Classification}
In the polarization task, PriSAR achieves an F1-score of 0.933, compared with 0.665 for the Real Image baseline and 0.763 for Flux. This result suggests that the generated samples contain useful polarization-related cues for downstream recognition.

\subsubsection{Azimuth Classification}
The clearest advantage of our method appears in the fine-grained azimuth estimation task, which remains difficult for generic text-to-image models. As shown in Table \ref{tab2}, models relying mainly on text prompts, such as SDXL and PixArt, achieve F1-scores below 0.45, while Flux reaches 0.769. In contrast, PriSAR reaches an F1-score of 0.939, indicating that the integration of 3D geometric priors provides a more informative viewpoint cue under the present setting.

\begin{figure}[htpb]
    \centering
    \includegraphics[width=0.88\textwidth]{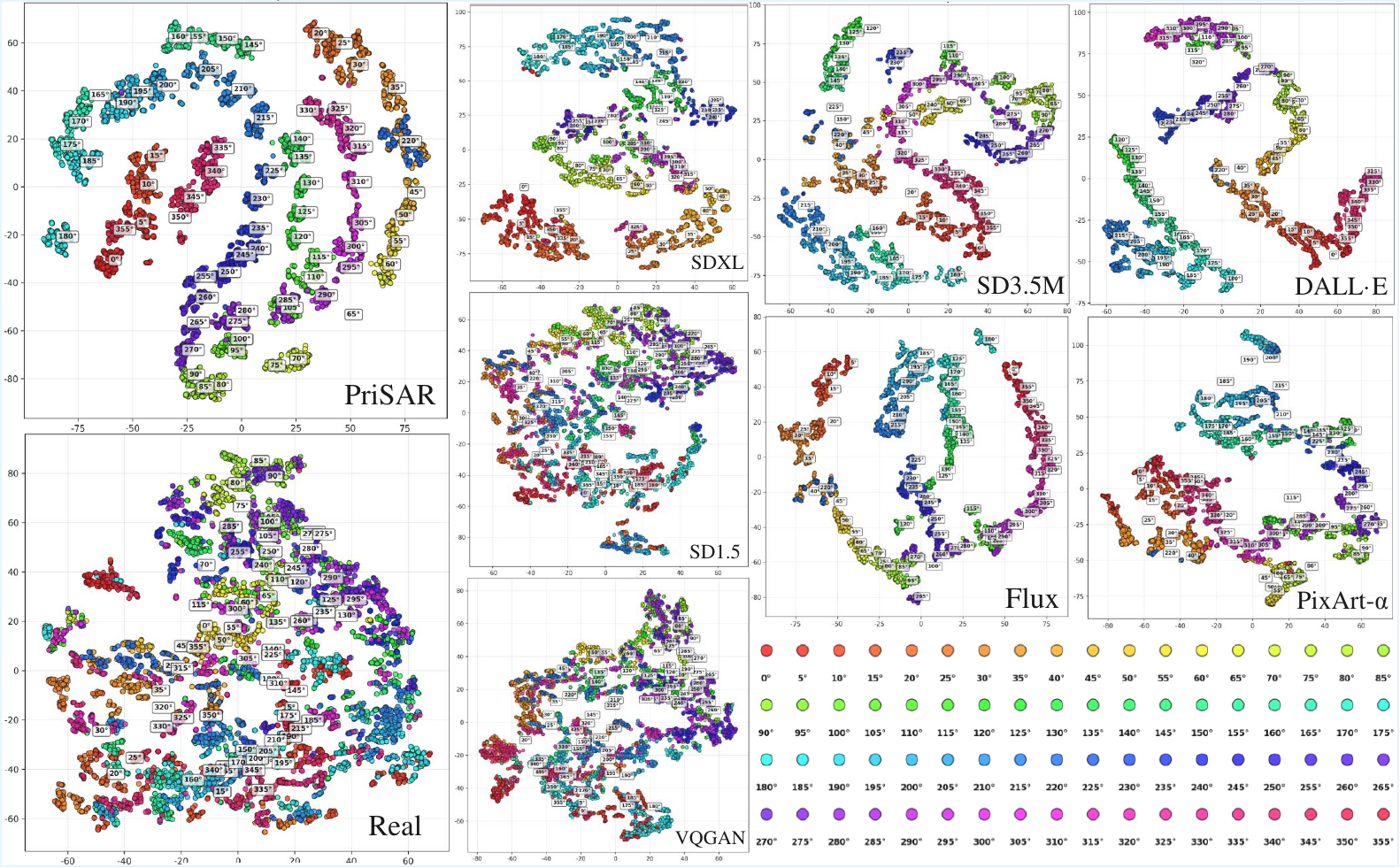} 
    \caption{T-SNE visualization of feature distributions in the azimuth estimation task. PriSAR forms clearer viewpoint-related clusters than the compared baselines.}
    \label{fig:tsne}
\end{figure}

Given the clear performance gap in the azimuth estimation task, we further use t-SNE to visualize the feature distributions of the generated images. Figure \ref{fig:tsne} compares the feature embeddings of PriSAR with representative baseline models. The feature manifolds of the baseline models remain more entangled across azimuth angles, indicating that text-driven models struggle to distinguish precise viewing directions. By contrast, PriSAR produces more separable clusters, suggesting that the injected 3D geometric priors help separate orientation-related features.

\begin{figure}[htpb]
    \centering
    \includegraphics[width=0.90\textwidth]{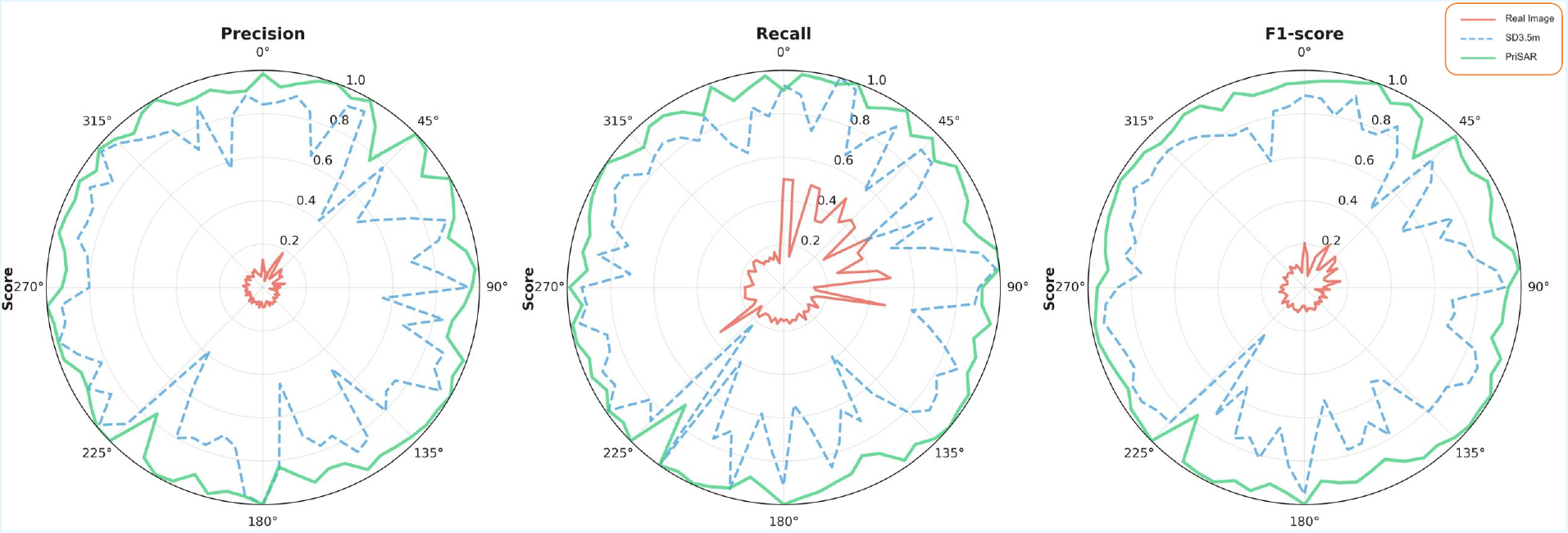} 
    \caption{Fine-grained ablation over the full $360^{\circ}$ azimuth range. PriSAR exhibits smoother Precision, Recall, and F1 trends, indicating stronger viewpoint consistency than the real-data baseline and vanilla SD3.5m.}
    \label{fig:polar}
\end{figure}

To conduct a fine-grained ablation study on the efficacy of the proposed 3D geometric prior, we quantitatively evaluated the azimuth estimation performance across the entire $360^{\circ}$ spectrum. Figure \ref{fig:polar} presents the polar plots detailing the Precision, Recall, and F1-score for three experimental settings.

As illustrated, the Real Image baseline forms a compact inner cluster near the origin, indicating limited generalization under sparse real-data training. The vanilla SD3.5m baseline expands the performance range but still exhibits a noticeable sawtooth pattern, revealing sensitivity to specific viewing angles. In contrast, PriSAR forms a smoother outer ring, which suggests improved azimuth-related stability across viewpoints.

\section{Conclusion}

In this paper, we presented PriSAR, a geometry-guided generative framework designed to alleviate the data scarcity bottleneck in SAR image generation. By introducing 3D-model-derived priors into a diffusion backbone, the framework aims to support controllability for observation-dependent SAR synthesis while remaining computationally more tractable than full electromagnetic simulation.

Our method employs a ray-casting-based point cloud simulator to capture view-dependent scattering cues and injects these geometric priors through a multi-modal fusion network, while LoRA enables efficient adaptation of Stable Diffusion 3.5 Medium to the SAR domain. Experiments on the real SAR aircraft dataset indicate consistent gains in image quality and azimuth-related controllability relative to strong baselines. The downstream augmentation results further suggest that the generated samples can be useful for training recognition models under sparse-angle settings. At the same time, we emphasize that the proposed prior branch should be understood as a lightweight geometry-guidance mechanism rather than a full electromagnetic solver. Nevertheless, the current validation is concentrated on a self-collected SAR aircraft dataset and on the sparse-angle completion setting, so broader evaluation on additional public benchmarks and acquisition conditions remains future work.

In future work, we aim to extend this geometry-guided framework to more complex scenarios, such as large-scale background clutter and multi-target interactions, to further broaden its applicability in remote sensing tasks.

\section*{Acknowledgments}
This work was supported by the National Natural Science Foundation of China [Grant Nos. 62201027 and 62271034].

\section*{CRediT authorship contribution statement}
Fan Zhang: Conceptualization, Supervision, Formal analysis, Funding acquisition, Project administration. Xuanting Wu: Software, Investigation, Methodology, Validation, Writing - original draft. Fei Ma: Conceptualization, Supervision, Funding acquisition, Project administration, Writing - review \& editing. Qiang Yin: Investigation, Resources, Validation. Yuxin Hu: Data curation, Investigation, Resources.

\section*{Declaration of competing interest}
The authors declare that they have no known competing financial interests or personal relationships that could have appeared to influence the work reported in this paper.



\FloatBarrier


\begin{thebibliography}{99}

\bibitem[Arjovsky et al.(2017)]{21} Arjovsky, M., Chintala, S., Bottou, L., 2017. Wasserstein GAN. Proc. Int. Conf. Mach. Learn. (ICML), 214--223.

\bibitem[Auer et al.(2009)]{13} Auer, S., Hinz, S., Bamler, R., 2009. Ray tracing simulation techniques for understanding high-resolution SAR images. IEEE Trans. Geosci. Remote Sens. 48, 1445--1456.

\bibitem[Balz(2006)]{15} Balz, T., 2006. Real-time SAR simulation of complex scenes using programmable graphics processing units. Proc. EUSAR, 1--4.

\bibitem[Batzolis et al.(2021)]{29} Batzolis, G., Stanczuk, J., Schonlieb, C.-B., et al., 2021. Conditional image generation with score-based diffusion models. arXiv:2111.13606.

\bibitem[Black Forest Labs(2024)]{48} Black Forest Labs, 2024. FLUX.1 [software]. \url{https://bfl.ai/}. (Accessed 21 April 2026).

\bibitem[Chen et al.(2025)]{9} Chen, H., Zhao, W., Zhang, R., Li, N., Li, D., 2025. Multiple object tracking in video SAR: A benchmark and tracking baseline. IEEE Geosci. Remote Sens. Lett. 22, Art. no. 4010905. doi: 10.1109/LGRS.2025.3592711.

\bibitem[Chen et al.(2024)]{46} Chen, J., Yu, J., Ge, C., et al., 2024. PixArt-alpha: Fast training of diffusion transformer for photorealistic text-to-image synthesis. Proc. Int. Conf. Learn. Represent. (ICLR).

\bibitem[Chen et al.(2016)]{5} Chen, S., Wang, H., Xu, F., et al., 2016. Target classification in synthetic aperture radar images using a convolutional neural network. IEEE Geosci. Remote Sens. Lett. 13, 156--160.

\bibitem[Dhariwal and Nichol(2021)]{26} Dhariwal, P., Nichol, A., 2021. Diffusion models beat GANs on image synthesis. Proc. Adv. Neural Inf. Process. Syst. (NeurIPS).

\bibitem[Esser et al.(2024)]{40} Esser, P., Kulal, S., Lang, A., et al., 2024. Scaling rectified flow transformers for high-resolution image synthesis. Proc. Int. Conf. Mach. Learn. (ICML).

\bibitem[Esser et al.(2021)]{45} Esser, P., Rombach, R., Ommer, B., 2021. Taming transformers for high-resolution image synthesis. Proc. IEEE CVPR, 12873--12883.

\bibitem[Franceschetti et al.(2003)]{14} Franceschetti, G., Migliaccio, M., Riccio, D., et al., 2003. SAR raw signal simulation of complex scenes. IEEE Trans. Aerosp. Electron. Syst. 39, 323--331.

\bibitem[Gao et al.(2025)]{12} Gao, F., Li, M., Wang, J., Sun, J., Hussain, A., Zhou, H., 2025. General sparse adversarial attack method for SAR images based on keypoints. IEEE Trans. Aerosp. Electron. Syst. 61, 14943--14960. doi: 10.1109/TAES.2025.3588821.

\bibitem[Goodfellow et al.(2014)]{17} Goodfellow, I., Pouget-Abadie, J., Mirza, M., et al., 2014. Generative adversarial nets. Proc. Adv. Neural Inf. Process. Syst. (NeurIPS).

\bibitem[Guo et al.(2017)]{20} Guo, J., Xu, B., Ju, S., 2017. Synthetic aperture radar image synthesis by using generative adversarial nets. IEEE Geosci. Remote Sens. Lett. 14, 1111--1115.

\bibitem[Ho et al.(2020)]{24} Ho, J., Jain, A., Abbeel, P., 2020. Denoising diffusion probabilistic models. Proc. Adv. Neural Inf. Process. Syst. (NeurIPS) 33, 6840--6851.

\bibitem[Hu et al.(2022)]{39} Hu, E.J., Shen, Y., Wallis, P., et al., 2022. LoRA: Low-rank adaptation of large language models. Proc. Int. Conf. Learn. Represent. (ICLR).

\bibitem[Huang et al.(2025)]{49} Huang, Z., Zhuang, Y., Zhong, Z., et al., 2025. X-Fake: Juggling utility evaluation and explanation of simulated SAR images. IEEE Trans. Image Process. 34, 7830--7844. doi: 10.1109/TIP.2025.3634988.

\bibitem[Huang et al.(2024)]{50} Huang, Z., Zhang, X., Tang, Z., Xu, F., Datcu, M., Han, J., 2024. Generative artificial intelligence meets synthetic aperture radar: A survey. IEEE Geosci. Remote Sens. Mag., early access. doi: 10.1109/MGRS.2024.3483459.

\bibitem[Isola et al.(2017)]{18} Isola, P., Zhu, J.-Y., Zhou, T., et al., 2017. Image-to-image translation with conditional adversarial networks. Proc. IEEE CVPR, 1125--1134.

\bibitem[Lewis et al.(2018)]{16} Lewis, B., Liu, J., Wong, A., 2018. Generative adversarial networks for SAR image realism. Proc. SPIE 10647, Algorithms for Synthetic Aperture Radar Imagery XXV, 1064709.

\bibitem[Ma et al.(2022a)]{6} Ma, F., Zhang, F., Xiang, D., Yin, Q., Zhou, Y., 2022a. Fast task-specific region merging for SAR image segmentation. IEEE Trans. Geosci. Remote Sens. 60, Art. no. 5222316. doi: 10.1109/TGRS.2022.3141125.

\bibitem[Ma et al.(2022b)]{7} Ma, F., Zhang, F., Yin, Q., Xiang, D., Zhou, Y., 2022b. Fast SAR image segmentation with deep task-specific superpixel sampling and soft graph convolution. IEEE Trans. Geosci. Remote Sens. 60, Art. no. 5214116. doi: 10.1109/TGRS.2021.3108585.

\bibitem[Malmgren-Hansen et al.(2017)]{36} Malmgren-Hansen, D., Kusk, A., Dall, J., et al., 2017. Improving SAR automatic target recognition using simulated data. Proc. IEEE Radar Conf., 1150--1153.

\bibitem[Moreira et al.(2013)]{1} Moreira, A., Prats-Iraola, P., Younis, M., et al., 2013. A tutorial on synthetic aperture radar. IEEE Geosci. Remote Sens. Mag. 1, 6--43.

\bibitem[Morgan(2015)]{3} Morgan, D.A.E., 2015. Deep convolutional neural networks for ATR from SAR imagery. Proc. SPIE 9475, Algorithms for Synthetic Aperture Radar Imagery XXII, 94750F. doi: 10.1117/12.2176558.

\bibitem[Mou et al.(2023)]{35} Mou, C., Wang, X., Xie, L., et al., 2023. T2I-Adapter: Learning adapters to dig out more controllable ability for text-to-image diffusion models. Proc. IEEE ICCV, 4296--4305.

\bibitem[Oliver and Quegan(2004)]{2} Oliver, C., Quegan, S., 2004. Understanding synthetic aperture radar images. SciTech Publishing, Raleigh, NC.

\bibitem[Pei et al.(2018)]{4} Pei, J., Huang, Y., Huo, W., et al., 2018. SAR automatic target recognition based on multiview deep learning framework. IEEE Trans. Geosci. Remote Sens. 56, 2196--2210.

\bibitem[Perera et al.(2023)]{28} Perera, M., Noshadi, A., Fookes, C., 2023. SAR despeckling using a denoising diffusion probabilistic model. IEEE Geosci. Remote Sens. Lett. 20, 1--5.

\bibitem[Perez et al.(2018)]{38} Perez, E., Strub, F., De Vries, H., et al., 2018. FiLM: Visual reasoning with a general conditioning layer. Proc. AAAI Conf. Artif. Intell. 32, 3942--3951. doi: 10.1609/aaai.v32i1.11671.

\bibitem[Podell et al.(2024)]{47} Podell, D., English, Z., Lacey, K., et al., 2024. SDXL: Improving latent diffusion models for high-resolution image synthesis. Proc. Int. Conf. Learn. Represent. (ICLR).

\bibitem[Radford et al.(2021)]{42} Radford, A., Kim, J.W., Hallacy, C., et al., 2021. Learning transferable visual models from natural language supervision. Proc. Int. Conf. Mach. Learn. (ICML), 8748--8763.

\bibitem[Raffel et al.(2020)]{43} Raffel, C., Shazeer, N., Roberts, A., et al., 2020. Exploring the limits of transfer learning with a unified text-to-text transformer. J. Mach. Learn. Res. 21, 1--67.

\bibitem[Ramesh et al.(2021)]{23} Ramesh, A., Pavlov, M., Goh, G., et al., 2021. Zero-shot text-to-image generation. Proc. Int. Conf. Mach. Learn. (ICML), 8821--8831.

\bibitem[Rombach et al.(2022)]{30} Rombach, R., Blattmann, A., Lorenz, D., et al., 2022. High-resolution image synthesis with latent diffusion models. Proc. IEEE CVPR, 10684--10695.

\bibitem[Rostami et al.(2019)]{11} Rostami, M., Kolouri, S., Eaton, E., Kim, K., 2019. Deep transfer learning for few-shot SAR image classification. Remote Sens. 11, 1374. doi: 10.3390/RS11111374.

\bibitem[Schreiber and Gupta(2005)]{37} Schreiber, E., Gupta, I.J., 2005. Scattering center analysis of man-made targets. IEEE Trans. Antennas Propag. 53, 178--184.

\bibitem[Shen et al.(2025)]{32} Shen, B., Liu, T., Gao, G., Chen, H., Yang, J., 2025. A low-cost polarimetric radar system based on mechanical rotation and its signal processing. IEEE Trans. Aerosp. Electron. Syst. 61, 4744--4765. doi: 10.1109/TAES.2024.3507776.

\bibitem[Song et al.(2021)]{25} Song, J., Meng, C., Ermon, S., 2021. Denoising diffusion implicit models. Proc. Int. Conf. Learn. Represent. (ICLR).

\bibitem[Tang et al.(2024)]{31} Tang, D., Cao, X., Hou, X., Jiang, Z., Liu, J., Meng, D., 2024. CRS-Diff: Controllable remote sensing image generation with diffusion model. IEEE Trans. Geosci. Remote Sens. 62, 1--14. doi: 10.1109/TGRS.2024.3453414.

\bibitem[Van den Oord et al.(2017)]{22} Van den Oord, A., Vinyals, O., Kavukcuoglu, K., 2017. Neural discrete representation learning. Proc. Adv. Neural Inf. Process. Syst. (NeurIPS).

\bibitem[Wu et al.(2024)]{41} Wu, X.,aa Jiang, L., Wang, P.S., et al., 2024. Point Transformer V3: Simpler, faster, stronger. Proc. IEEE CVPR, 4840--4851.

\bibitem[Yang(2020)]{44} Yang, R., 2020. PyTorch Image Models Multi-Label Classification [software]. GitHub. \url{https://github.com/yang-ruixin/PyTorch-Image-Models-Multi-Label-Classification}. (Accessed 21 April 2026).

\bibitem[Zhang et al.(2019)]{10} Zhang, F., Zhou, Y., Mao, S., et al., 2019. Data augmentation for SAR target recognition using generative adversarial network. IEEE Geosci. Remote Sens. Lett. 16, 1467--1471.

\bibitem[Zhang et al.(2023)]{34} Zhang, L., Rao, A., Agrawala, M., 2023. Adding conditional control to text-to-image diffusion models. Proc. IEEE ICCV, 3836--3847.

\bibitem[Zhang et al.(2026)]{8} Zhang, T., Xie, N., Quan, S., Wang, W., Wei, F., Yu, W., 2026. Polarimetric SAR ship detection based on the sub-look decomposition technology. IEEE Trans. Radar Syst. 4, 35--49.

\bibitem[Zhang et al.(2025)]{51} Zhang, X., Zhuang, Y., Guo, Q., et al., 2025. Ph-GAN: Physics-inspired GAN for generating SAR images under limited data. Proc. IEEE/CVF ICCV, 29075--29085.

\bibitem[Zhong et al.(2025)]{33} Zhong, F., Gao, F., Liu, T., et al., 2025. Scattering characteristics guided network for ISAR space target component segmentation. IEEE Geosci. Remote Sens. Lett. 22, Art. no. 4009505. doi: 10.1109/LGRS.2025.3576662.

\bibitem[Zhu et al.(2017)]{19} Zhu, J.-Y., Park, T., Isola, P., et al., 2017. Unpaired image-to-image translation using cycle-consistent adversarial networks. Proc. IEEE ICCV, 2223--2232.

\end{thebibliography}
\end{document}